# Close relationship between superconductivity and the bosonic mode in $Ba_{0.6}K_{0.4}Fe_2As_2$ and $Na(Fe_{0.975}Co_{0.025})As$

Zhenyu Wang[1,2,§], Huan Yang[1,§], Delong Fang[1], Bing Shen[2], Qiang-Hua Wang[1], Lei Shan[2], Chenglin Zhang[3], Pengcheng Dai[2,3] and Hai-Hu Wen[1*]

[1] Center for Superconducting Physics and Materials, National Laboratory of Solid State Microstructures and Department of Physics, Nanjing University, Nanjing 210093, China
[2] National Laboratory for Superconductivity, Institute of Physics and National Laboratory for Condensed Matter Physics, Chinese Academy of Sciences, Beijing 100190, China
[3] Department of Physics and Astronomy, The University of Tennessee, Knoxville, Tennessee 37996-1200, USA
[§] These authors contributed equally to this work.

**Since the discovery of high temperature superconductivity in the iron pnictides and chalcogenides in early 2008, a central issue has been the microscopic origin of the superconducting pairing. Although previous experiments suggest that the pairing may be induced by exchanging the antiferromagnetic spin fluctuations and the superconducting order parameter has opposite signs in the electron and hole pockets as predicted by the S± pairing model, it remains unclear whether there is a bosonic mode from the tunneling spectrum which has a close and universal relationship with superconductivity as well as the spin excitation. In this paper, based on the measurements of scanning tunneling spectroscopy, we show the clear evidence of a bosonic mode with the energy identical to that of the neutron spin resonance in two completely different systems $Ba_{0.6}K_{0.4}Fe_2As_2$ and $Na(Fe_{0.975}Co_{0.025})As$ with different superconducting transition temperatures. In both samples, the superconducting coherence peaks and the mode feature vanish simultaneously inside the vortex core or above $T_c$, indicating a close relationship between superconductivity and the bosonic mode. Our data also demonstrate a universal ratio between the mode energy and superconducting transition temperature, that is $\Omega/k_BT_c \approx 4.3$, which underlines the unconventional mechanism of superconductivity in the iron pnictide superconductors.**

According to the BCS theory, the superconducting state is achieved by the quantum condensation of paired electrons, and the electron pairing is provided by the electron-phonon interaction. The compelling evidence for this model was the tunneling spectrum $dI/dV$ vs. $V$, on which some fine structures, due to the electron-phonon coupling, appear at the voltages of $\Delta+\Omega$ (or at $-\Delta-\Omega$) with $\Delta$ the superconducting gap and $\Omega$ the energy of a typical phonon mode[1-4]. Hence, one should be able to observe the clear dips (or peaks) in the second derivative of tunneling curve, i.e., $d^2I/dV^2$ vs. $V$ at energies of $\Delta+\Omega$ (or at $-\Delta-\Omega$). Reason for this is due to the energy dependent gap function $\Delta(\varepsilon)$, when taking the electron-phonon coupling into account, it shows the anomalies at $\Delta+\Omega$ (or at $-\Delta-\Omega$) and thus influences the quasi-particle density of states (DOS) which is proportional to $dI/dV$[1-4]. Inspired by this interesting scenario, it was found that similar "anomaly" appeared outside the gap in hole-doped cuprate $Bi_2Sr_2CaCu_2O_{8+\delta}$ measured by scanning tunneling microscopy (STM)[5], although it was argued that this peak-dip-hump feature might not be relevant for superconductivity, but resulted from inelastic tunneling through the insulating oxide layers[6] or simply from some Einstein mode of phonons[7]. For the electron-doped high-$T_C$ cuprate superconductors $Pr_{0.88}LaCe_{0.12}CuO_4$, STM measurements have revealed a bosonic excitation at an energy of 10.5 meV, which is consistent with the neutron spin resonance energy, thus indicate the spin fluctuation mediated superconductivity in this material[8-10]. In the iron pnictide/chalcogenide superconductors, one of the proposed pictures for the pairing is through exchanging the antiferromagtic (AF) spin fluctuations by the two electrons on the electron (hole) pockets, and they are scattered to the hole (electron) pockets leading to a pairing order parameter with opposite signs on the electron and hole pockets, this is termed as the S± pairing model[11,12]. Inelastic neutron scattering data have given some preliminary evidence for this pairing gap, namely a resonance peak appears at the momentum Q = (π, 0) (in the unfolded Brilloun zone)[13]. In addition, some evidence for supporting this model is inferred from the quasi-particle interference (QPI) pattern based on the spatial mapping of the local density of states (LDOS) in the STM measurements in Fe(Se,Te)[14]. Therefore it is interesting to see whether the tunneling spectrum $dI/dV$ vs. $V$ will exhibit some anomalies related to the neutron resonance or the unique S± pairing. Recently, a dip-hump feature in the scanning tunneling spectroscopy (STS) of an optimally hole-doped $BaFe_2As_2$ is observed, and the mode energy seems to be close to the neutron spin resonance energy[15]. However, it is still possible that the mode is induced by the interaction between the electrons and a phonon mode. Here we present evidence for this bosonic mode in two totally different systems with different superconducting transition temperatures ($T_c$). These data suggest a close relationship between superconductivity and the bosonic mode associated with spin excitations instead of phonons.

Figure **1a** shows the topographic image of the $Ba_{0.6}K_{0.4}Fe_2As_2$ sample measured by our STM with the tunneling current $I_t$ = 200 pA and bias voltage $V$ = 40 mV. Atomically resolved structure is clearly seen. The inset to Fig. **1a** shows a closer view of the topography, the atomic structure turns out to be a square lattice with a constant of 5.4 Å with a mixture of white and dark spots on the image. By counting the numbers of the white and dark spots, we

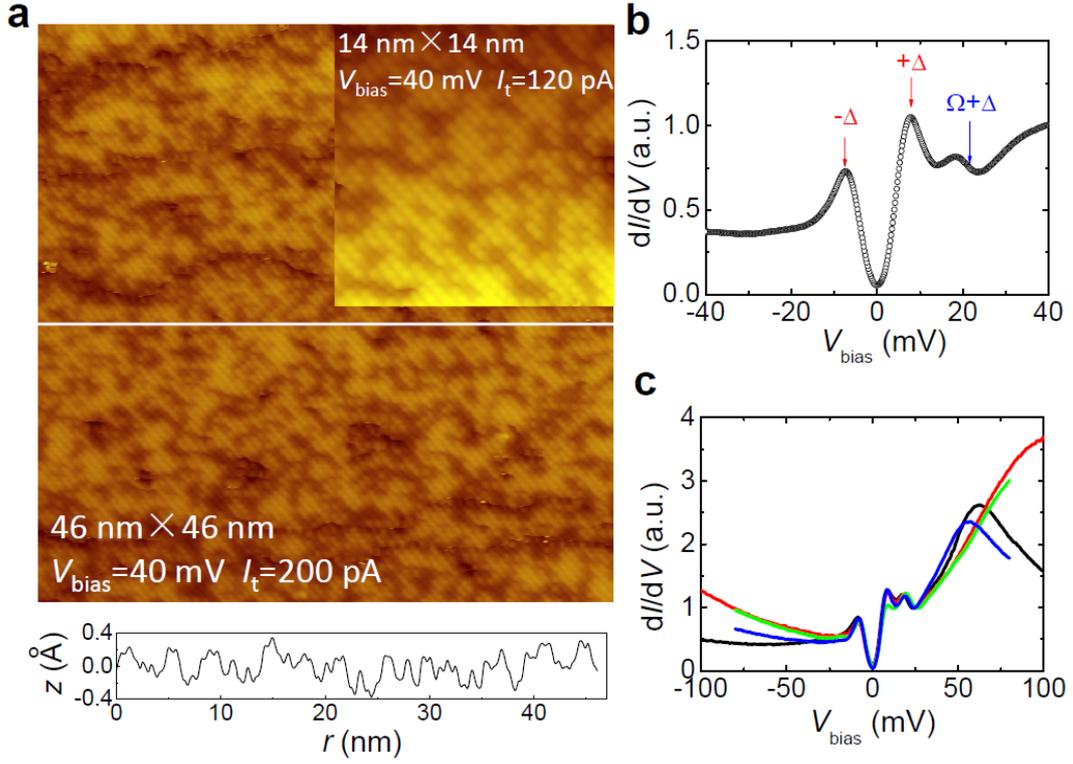

**Figure 1 | Topographic STM image and the tunneling spectra of the $Ba_{0.6}K_{0.4}Fe_2As_2$ single crystal. a**, STM image taken with a bias voltage of $V_{bias}$ = 40 mV and tunneling current of $I_t$ = 120 pA. An atomically resolved square lattice with a constant of 5.4 Å is observed. The inset of **a** shows a more closer view. The curve below **a** shows a height distribution measured along the white line marked in **a**. **b,** A typical tunneling spectrum measured at 1.7 K. The superconducting coherence peaks are marked by the arrows with "Δ" and "-Δ". The arrow with "Ω+Δ" reflects the position of the bosonic mode. A clear asymmetric spectrum is observed, which may be induced by the shallow bands close to the Fermi energy. **c**, Several typical tunneling spectra measured to a wider energy, exhibiting an hump at about 60 mV or higher energies.

find that the ratio is actually very close to the chemical doping ratio Ba/K = 6/4, indicating that the surface layer is constructed by the Ba and K atoms. The high quality of the sample is justified by the data in Fig. S1 of the Supplementary Information. Fig. 1b shows a typical tunneling spectrum measured on the surface. One can see several interesting features: (1) The tunneling spectrum exhibits two clear peaks at the energy levels of about ±7.3 meV which are associated with the superconducting gaps. (2) An additional peak appears at the energy of about 21.5±0.8 meV (determined at the middle of right-wing of the peak, addressed later). Apparently both the superconducting coherence peak and the peak at about 21.5±0.8 meV are more prominent on the right hand side. We determined the mode energy at the middle of the right wing of the peak because this was adopted in the conventional superconductors to determine the phonon energy, according to the Eliashberg theory[1-4]. For unconventional superconductors with sign-reversing gaps, the peak (as a consequence of the sign-reversing gap) position depends on the superconducting energy gap as well as the correlation effect[16-19]. We will come to this point later. (3) The spectrum is quite asymmetric. As shown in Fig. 1c, the DOS is generally low in the negative-bias side, and gets much higher at positive bias voltage. On some curves, a peak appears at about 60 mV, and it can even move to higher energies at different sites, but the feature of superconductivity in the low energy region shows no obvious change. This asymmetric feature can be well explained as result of the band edge effect.

Actually from the ARPES data, one can indeed see that several bands are close to the Fermi energy (within 100 meV)[20,21]. The peak with the characteristic energy of about 60 mV or higher may be induced by the accumulation of DOS of both the electron and hole bands. Our simulation tells that indeed this band edge effect can strongly enhance the DOS as well as the related features in the positive-bias side (see later).

In order to know whether this mode is closely related to superconductivity, we investigate the temperature and magnetic field dependence of the tunneling spectra. In Fig. 2a, we show the evolution of the STS with temperatures up to 40 K, which is just above $T_c$. The black solid lines shown together represent those by doing the Fermi-Dirac convolution on the data at 1.7 K[22]. It is clear that at 40 K the mode feature is getting almost invisible compared to the convoluted data of 1.7 K. Therefore both the superconducting coherence peak and the mode are suppressed and smeared out with increasing temperature and finally disappeared above $T_c$. To have a comprehensive understanding on the data, we normalize the $dI/dV$ vs. $V$ spectrum measured at different temperatures by the one measured at 40 K and show the results in Fig. 2b. Shown here together are theoretical curves (colour lines) calculated using a two anisotropic s-wave gaps based on the Dynes model[23]. Actually the calculation with one single anisotropic gap can also fit the data. Both ways yield a dominant gap of 7.3±0.5 meV which is close to our previous results in the STM measurements in the same kind of samples[24]. The details about the fitting is given in the Supplementary Information. To illustrate the relationship between superconductivity and the bosonic mode, we determine the area of both the coherence peak and the mode under the tunneling spectrum. It is known that the tunneling spectrum would show up as the red line if the electron-boson coupling have no influence, therefore we determine the weight corresponding to superconductivity by integrating the area under the coherence peak, as marked by the red colour in Fig. 2**c**. For the bosonic mode, we calculate the area between the mode peak and the calculated curve, as marked by the blue colour in Fig. 2**c**. The results arising from this calculation about the superconducting coherence peak and the bosonic mode are shown in Fig. 2**d**. It is evident that both will vanish at $T_c$, indicating a close relationship between superconductivity and the bosonic mode.

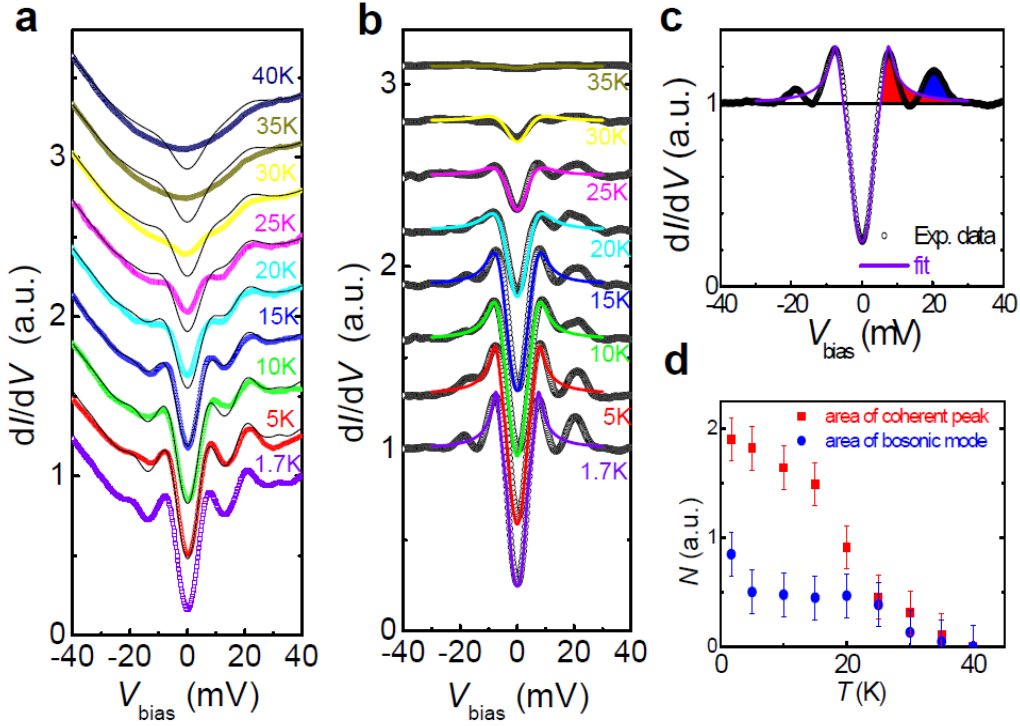

**Figure 2 | Scanning tunneling spectrum and their temperature dependence on the $Ba_{0.6}K_{0.4}Fe_2As_2$ single crystal. a,** The evolution of the STS spectra with temperature increased from 1.7 K to 40 K. One can see that both the coherence peak and mode peak are suppressed and gradually vanish above $T_c$. The solid lines represent the data at 1.7 K convoluted with the Fermi-Dirac function. **b,** The STS normalized by the one measured in normal state (at 40 K). The symbols represent the experimental data, the colour lines are the theoretical fit to the data with the Dynes model with two anisotropic gaps. **c,** The normalized spectrum measured at 1.7 K, the red and the blue areas are those calculated for showing the weight of the coherence peak and the bosonic mode. **d,** The temperature dependence of the area for the coherence peak and the mode, as marked by the red and blue areas respectively in **c**. Clearly they both vanish at about $T_c$. The error bars in **d** indicate the error in the fitting process using Eq. S1.

An alternative way to check the relationship between superconductivity and the bosonic mode is to suppress the superconductivity by a high magnetic field and to see how the mode changes inside the vortex core. For this purpose, we applied a magnetic field (11T) during the measurements. By measuring the 2D spatial distribution of the LDOS at zero energy, we get a vortex pattern with a slightly distorted triangular lattice. In Fig. 3**a**, we show the image of the vortices with a high resolution. It seems that the vortex structure is not as disorder as in the $Ba(Fe_{1-x}Co_x)_2As_2$, this is perhaps due to the fact that the doped atoms go directly to the Fe sites leading to larger disorder in the Co-doped samples[25]. With the well measured vortex pattern, we can easily locate and do the careful measurements of the STS curve by crossing one single vortex. By following the trace as marked by the white line in Fig.3**a**, we measured the spatially resolved spectra ranging from the outside to inside of the vortex core. A typical set of data are presented in Fig. 3**b**. It is clear that, outside the vortex core, the spectrum looks quite similar to that measured at zero magnetic field: a very weak suppression occurs to the coherence peak and the mode feature when the field is applied. However, both peaks are strongly suppressed and even cannot be distinguished one from another when it is inside the vortex core. In Fig. 3**c**, we present spatial evolution of the spectra measured from outside to inside of the vortex cores with a step of 3.6 Å. One can see that both the coherence peak and the mode peak are suppressed simultaneously and finally they are absent by forming a very broad and round hump. In order to resolve the evolution more clearly, we present in Fig. 3**d** the second derivative curves $d^2I/dV^2$ vs. $V$. A strong "dip" appears at about 21.5±0.8 meV, as marked by the dashed line. If we attribute this strong "dip" to the electron-boson interaction, being analogous to that in the phonon mediated superconductors, the bosonic mode energy here is about $\Omega$ = 21.5-7.3 = 14.2 meV. This value is very close to 14-15 meV, the energy of the neutron resonance obtained from the same sample[26]. We should mention that either the kink on the $dI/dV$ vs. $V$ curve or the "dip" on the $d^2I/dV^2$ vs. $V$ curve at 21.5±0.8 meV do not change the position, seemingly contradicting with a weakened pairing strength when going from outside to inside of the vortex core. The apparent gapped-feature at the center of the vortex core is very unusual. Similar phenomenon was observed in the cuprates and explained as an effect of a competing order[27,28]. In our present systems, further systematic work are certainly desired to solve this issue. Another interesting phenomena is that in our present study, we didn't see the in-core Andreev bound state peak, as we observed in previous experiments[29]. This could be due to several reasons: (1) In present experiment, the coherence peak arises mainly from the larger gap (7-8 meV), while the previous STS curve that exhibits the in-core Andreev bound states mainly comes from the small gap (3-4 meV). According to the extended theory of the S± model, the DOS and the gap value of each Fermi surface may have the relation $N_S/N_L = (\Delta_L/\Delta_S)^2$, therefore the DOS ($N_s$) on the Fermi surface of the small gap is much larger than that of the larger gap ($N_L$), thus the in-gap Andreev bound states are mainly arising from the DOS of the small gap[30]. (2) The distinction could be due to the different surface states in these two cases. We must emphasize that, the bulk qualities of the two samples are exactly the same, both are very good.

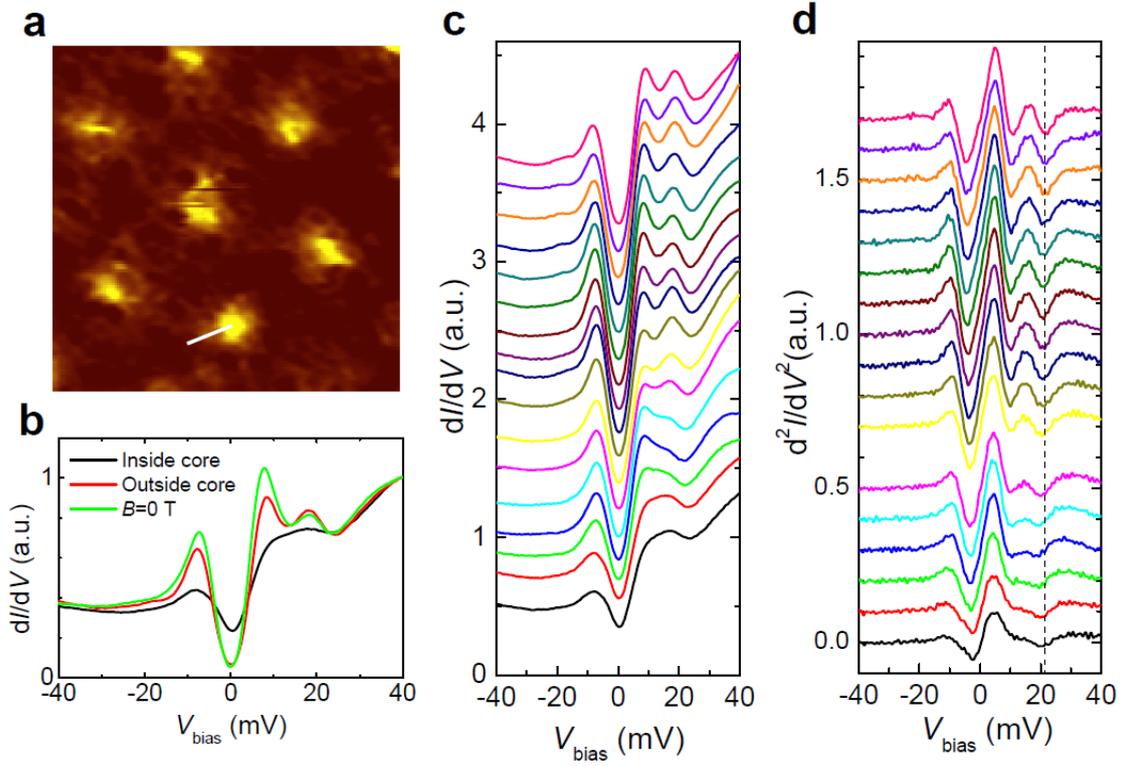

**Figure 3 | Spatial mapping of the LDOS and STS across a vortex core on the $Ba_{0.6}K_{0.4}Fe_2As_2$ single crystal. a,** A high resolution topographic image (39 nm × 39 nm) of the vortex lattice measured by the LDOS at zero bias. **b,** The Tunneling spectra measured at zero magnetic field (green line), outside the vortex core (red line) and inside the vortex core (dark line) when a 11 T magnetic field is applied. **c,** The spatially resolved tunneling spectrum $dI/dV$ vs. $V$ measured from outside (upper curve) to the center of the vortex core (bottom) with a step of 3.6 Å. No in-core Andreev bound state peaks are observed. Both the coherence peak and the mode peak are suppressed and gradually disappeared inside the vortex core, showing the close relationship between the superconducting coherence peak and the peak of the bosonic mode. **d,** The second derivative curve $d^2I/dV^2$ vs. $V$. The bosonic mode appears here at about 21.5 mV as a sharp "dip". The "dip" feature is gradually smeared up inside the vortex core.

Although we have shown that there is a strong mode feature outside the superconducting gap of $Ba_{0.6}K_{0.4}Fe_2As_2$, and the mode has a similar energy as the neutron resonance energy, one can still argue that it could be induced by the interaction between the quasiparticles and the Einstein phonon mode which probably has an energy, by accident, to be around 14 meV. In order to show that this is not the case, we select a totally different system with different $T_c$, that is, $Na(Fe_{0.975}Co_{0.025})As$ ($T_c$ = 21 K). This system is perfect for the STM measurement since we can get an unpolar surface with well ordered Na atoms[22,31,32]. In Fig.4**a** we present the topography of the cleaved surface on this 111 sample with a tunneling current of $I_t$=200 pA and voltage of $V$ = 40 meV. Beside the atomic lattice, one can see a block with the rectangular shape in the morphology. This was argued to correspond to one single Co impurity and discussed elsewhere[32]. Here we show the tunneling spectrum along the line marked with blue dots in Fig.4**a** with every half lattice constant. The spatial evolution of the STS is shown in Fig.4**c**. One can see that the STS exhibit a flat zero bottom, indicating a full superconducting gap. Moreover, the superconducting coherence peaks are very clear and strong here, yielding a gap value of about 5.5 meV. This gap value is quite close to that observed by the angle-resolved-photoemission spectroscopy (ARPES) which indicates an identical gaps on both the hole and electron pockets[33]. In addition, just outside the superconducting coherence peak, a little hump appears at the energy of about 13.3±0.8 meV on both sides. In order to show this more clearly, we present the second derivative curves $d^2I/dV^2$ in Fig.4**d**. One can see a clear dip at the energy of about 13.3±0.8 meV. If we still follow the way that used for $Ba_{0.6}K_{0.4}Fe_2As_2$ to determine the mode energy for $Na(Fe_{0.975}Co_{0.025})As$, that would give $\Omega$ = 13.3-5.5 ≈ 7.8 meV. In Fig.4**b**, we show the difference of the imaginary part of the spin susceptibility measured at 5 and 25 K. One can see that a resonance peak appears at around 7 meV at Q = *(1/2,1/2,2)*. This resonance energy once again is very close to the mode energy determined from the tunneling measurement. In Fig.5a, we show the temperature dependence of the tunneling spectrum, one sees again both the superconducting coherence peak and the mode at 1.7 K. With increasing temperature, both features are weakened simultaneously and vanish above T$_c$. In all the STS curves, we see a clear asymmetry. This asymmetry extends even to a temperature well above T$_c$. This asymmetry, as we see in the Ba$_{0.6}$K$_{0.4}$Fe$_2$As$_2$, is argued to be the result of the band edge effect. Interestingly, as shown in Fig.2a, when there is a weak anomaly feature on the positive bias side in the normal state, the superconducting coherence peak and the mode feature are enhanced also on the same side. By taking the data at 25 K (> T$_c$ = 21 K) as the background and divide it for the STS at all other temperatures, we get STS with a more symmetric shape and show them in Fig.5b. After doing this, the mode feature becomes very clear. By applying a magnetic field of 11 T at 1.7 K, we measured the spatial distribution of the LDOS at zero energy. The data is shown in Fig.5c, one can see a slightly distorted triangle vortex lattice again in this sample. In Fig.5d, we show the spatial evolution of the STS measured along the arrowed white-line indicated in Fig.5c. One sees a very similar feature as observed in Ba$_{0.6}$K$_{0.4}$Fe$_2$As$_2$. It is clear that both the superconducting coherence peak and the mode feature observed outside the gap are suppressed simultaneously. It is interesting to remark that the STS measured at the vortex core looks quite similar to that measured above T$_c$. This asymmetric background both at the vortex core center and above T$_c$ is also observed in LiFeAs[22,31], and is naturally explained as due to the shallow band edge effect.

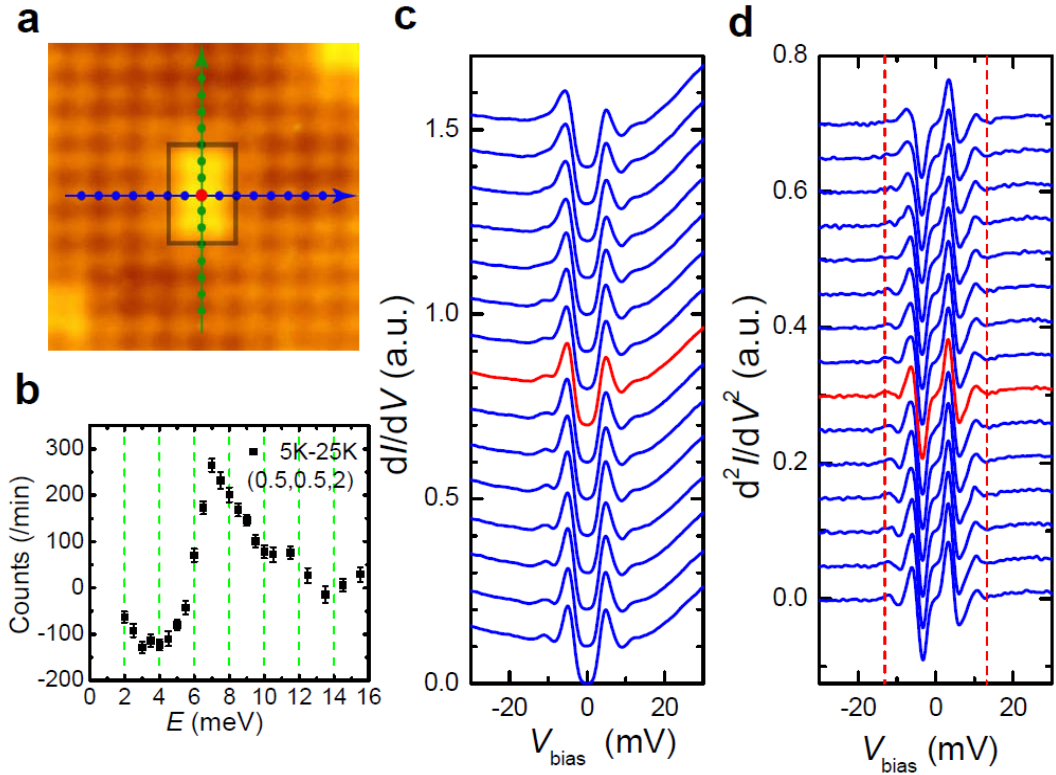

**Figure 4 | Spatially resolved STS on Na(Fe$_{0.975}$Co$_{0.025}$)As single crystal. a,** Spatially resolved atomic lattice with a constant of 3.80 Å measured with a tunneling current of I$_t$=200 pA and voltage of V = 40 meV. A rectangular shape with 2×3 atoms is observed, which is well associated with a single Co impurity in the middle and will be presented elsewhere. **b,** The difference of the imaginary part of the spin susceptibility measured at 5 and 25 K. One can see that a resonance peak appears at around 7 meV at Q = (1/2,1/2,2). **c,** The Tunneling spectra measured at zero magnetic field at the positions marked by the blue dots in **a**. The red line represents the one detected at the center of the 2×3 rectangular block, supposed to be the site of the Co impurity. **d,** The second derivative curve $d^2I/dV^2$ vs. *V*. The bosonic mode is marked by the dashed line, where a "dip" appears at about 13.5 mV. The error bars in **b** indicate statistical errors of one standard deviation.

The coincidence of the bosonic mode energy and the neutron resonance energy in two completely different systems is quite intriguing and indicates that the mode feature is unlikely induced by phonons. Given the dramatic differences in crystal structures between Ba$_{0.6}$K$_{0.4}$Fe$_2$As$_2$ and Na(Fe$_{0.975}$Co$_{0.025}$), it is difficult to imagine that two phonons would occur at exactly the energy of the resonance for the two systems. Therefore, our results suggest that the electrons are strongly coupled to the spin excitations. By taking account of the ratio between $\Omega$ and T$_c$, there is a universal ratio in the two different samples: $\Omega/k_BT_c$ = 14.2×11.6/38 = 4.33±0.25 in Ba$_{0.6}$K$_{0.4}$Fe$_2$As$_2$ and 7.8×11.6/21 = 4.31±0.44 in Na(Fe$_{0.975}$Co$_{0.025}$)As (taking ±0.8 meV as the error bar of $\Omega$). In the optimally doped Ba(Fe$_{1-x}$Co$_x$)$_2$As$_2$, since no STM data have been reported so far about the bosonic mode, we take the value of the neutron resonance energy, which is about 9.5 meV[34], using T$_c$ = 25 K, we have $E_r/k_BT_c = 4.28$. Meanwhile it is concluded that there is a universal ratio[35] between the resonance energy E$_r$ and T$_c$, which is about 4.6 in many kinds of the iron pnictide superconductors. In Fig. 6 we present the resonance energy derived from the neutron scattering experiments by the open symbols[13,26,34-43] for different systems, versus the superconducting transition temperatures. We also plot the mode energy determined from our tunneling measurements. Remarkably, our data points fall quite well onto the so-called universal plot.

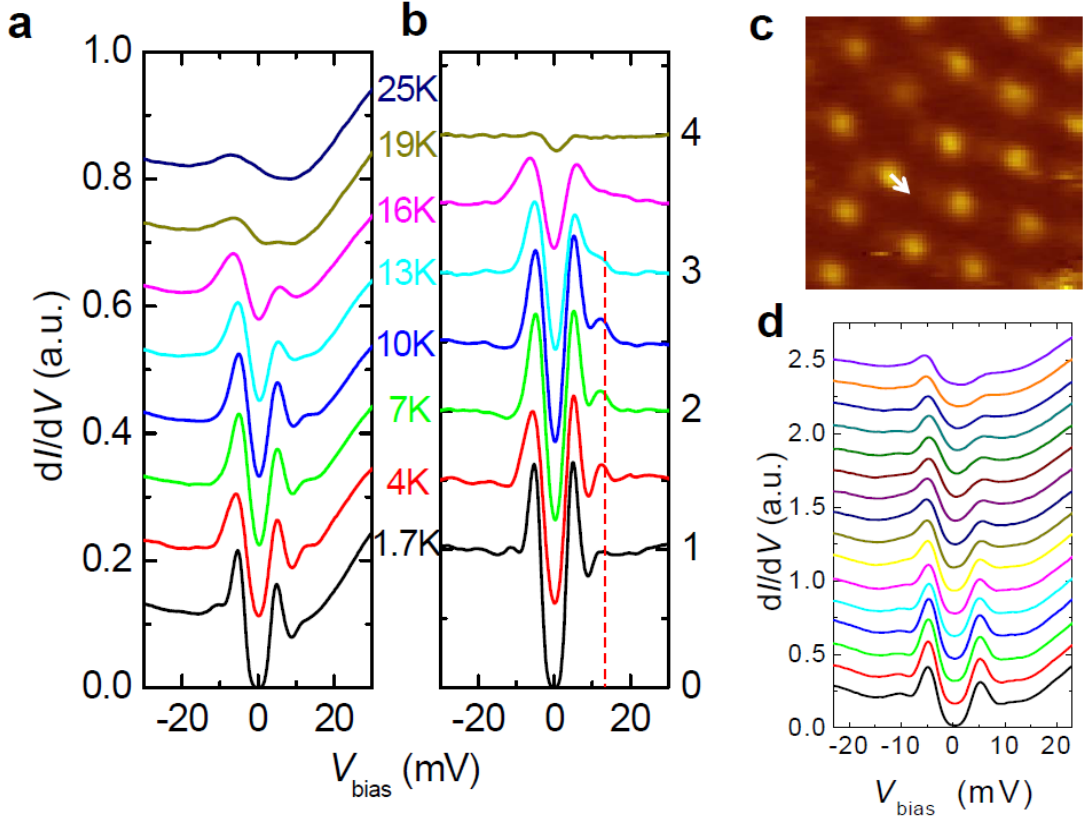

**Figure 5 | Temperature dependence of the tunneling spectra on the Na(Fe$_{0.975}$Co$_{0.025}$)As single crystal. a,** The evolution of the STS spectra with temperature increased from 1.7 K to 25 K. One can see a small hump appears above the superconducting coherence peak. The all spectra are asymmetric with an enhancement on the left hand side, which is attributed to the band edge effect. **b,** The STS normalized by the one measured in normal state (at 25 K). **c,** A high resolution topographic image (39 nm × 39 nm) of the vortex lattice measured by the LDOS at zero bias at a magnetic field of 11 T. One can see a slightly distorted vortex lattice structure. **d,** Spatially resolved tunneling spectra $dI/dV$ vs. $V$ measured from outside (bottom curve) to the center of the vortex core (upper) with a step of 3.6 Å.

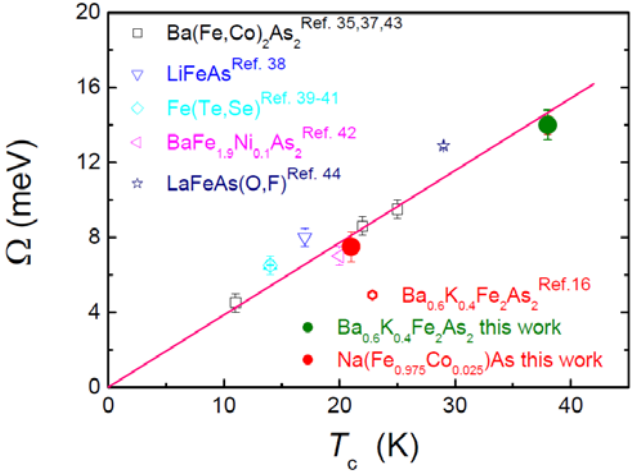

**Figure 6 | An universal relation between the mode energy and the superconducting transition temperature $T_c$.** The open points with error bars represent the neutron resonance energies versus $T_c$ for many different samples, the filled dark green and red circles show the mode energy $\Omega$ versus $T_c$ of our tunneling experiments on the Ba$_{0.6}$K$_{0.4}$Fe$_2$As$_2$ and Na(Fe$_{0.975}$Co$_{0.05}$)As single crystals, respectively. The red line shows a universal relation: $(E_r,\Omega)/k_BT_c$ = 4.4. The error bars of the data in this work indicate standard deviation of Gaussian distribution fitting of $\Omega$, those for open symbols are taken from the literatures.

The close relationship between electron tunneling mode and neutron spin resonance in two different systems strongly suggest that the pairing should have a magnetic origin. In this context, there are two scenarios in explaining the neutron resonance and the mode feature in tunneling measurements: (1) It is the consequence of the sign-reversal of the gap; (2) It is this mode that mediates the pairing, namely it acts as the pairing glue. Now we discuss on these two possibilities. A general argument against the idea of "pairing glue" for the neutron spin resonance stems from two concerns: (1) the absolute weight of the neutron resonance mode is too weak to account for the superconducting condensation in most unconventional superconductors[17], although in optimally doped YBa$_2$Cu$_3$O$_{7-\delta}$ the mode appears to have sufficient weight to account for the superconducting condensation energy[44]; and (2) more seriously, the mode vanishes above $T_c$ in optimally doped superconductors which is not like the phonon that persists all the way crossing $T_c$. According to the Eliashberg theory on superconductors with a sign-reversing gap, a resonance mode due to the particle hole excitation will be formed within $2\Delta$[16-19]. In a general point of view, the mode energy may be written as $\Omega = 2\Delta[1-f(U/t)]$ with an unknown function $f(U/t)$ being less than 1 and depending on the ratio between the repulsive energy $U$ and the kinetic energy $t$. Assuming that $\Delta \approx 2 \sim 2.5 k_B T_c$ in the case of strong coupling, we indeed have $\Omega/k_BT_c \approx$ 4-5 if the correlation is mild ($f(U/t) << 1$). Theoretically, a precise expression about $f(U/t)$ is still lacking. In addition, by following this argument, as far as we know, there is no explicit way to prove that the mode feature should locate at the dip position of the second derivative curve $d^2I/dV^2$ vs. $V$, as observed in our data and the phonon mediated superconductors[1-4]. Therefore our results will stimulate further theoretical efforts to get a quantitative understanding.

The other scenario regards the bosonic mode as the pairing glue. In order to get a basic assessment on this model, and especially for understanding the asymmetry of the mode features with respect to the Fermi level, we performed self-consistent microscopic calculations using the four band (two electron and two hole bands) Eliashberg formalism. This type of theory was applied for the d-wave pairing in cuprates with the assumption that the electrons are coupled to resonance spin fluctuation mode revealed by inelastic neutron scattering[45-46]. Along a similar line, we assume that the observed (π, 0) neutron resonance mode (for example 14 meV in $Ba_{0.6}K_{0.4}Fe_2As_2$) mediates pairing in the present system. The detailed simulation process is presented in the Supplementary Information and the results are given in Fig.S3 and Fig.S4. It is found the theoretical simulation can capture the basic features of the tunneling spectra by properly choosing the fitting parameters. In particular, in this simulation, the mode feature is situated at the dip position of the second derivative curve $d^2I/dV^2$ vs. $V$. However, the strength of the mode feature at the energy $\Delta+\Omega$ simulated by the Eliashberg theory is much weaker than the experimental one shown in Fig. 2**b**. The fact that the tunneling mode feature disappears above $T_c$ and inside the vortex core, and its energy is at $\Delta+\Omega$, suggest that the peak should be strongly related to superconductivity. The universal ratio of $\Omega/k_BT_c \approx 4.3$ in two different systems with distinct transition temperatures and the coincidence with the neutron resonance support the picture of pairing through spin fluctuations, although the bosonic mode itself may be a consequence of the sign-reversing superconducting gap.

**Acknowledgements**

We acknowledge the useful discussions with A. V. Chubukov, S. Kivelson, M. Norman, I. I. Mazin, I. Eremin and A. Balatsky. We specially thank A. A. Golubov, J. Schmalian and I. I. Mazin for their preliminary calculations in helping us to understand the data based on the Eliashberg theory. This work was supported by NSF of China, the Ministry of Science and Technology of China (973 projects: 2011CBA00100, 2012CB821403, 2012CB21400, 2010CB923002, 2011CB922101) and PAPD. The single crystal growth effort at UTK is supported by the U.S. DOE BES No. DE-FG02-05ER46202.

**Author contributions**

The low-temperature STS measurements were finished by ZYW, HY and DLF, and helped by HHW and LS. The samples were prepared by BS and CLZ. The simulation based on the Eliashberg theory was finished by QHW. HHW coordinated the whole work and wrote the manuscript which was supplemented by QHW, ZYW and HY, revised by PD. All authors have discussed the results and the interpretation.

**Competing financial interests**

The authors declare that they have no competing financial interests.

* Correspondence and requests for materials should be addressed to Hai-Hu Wen at hhwen@nju.edu.cn.

# SUPPLEMENTARY INFORMATION

## I. Quantitative analysis of the $Ba_{0.6}K_{0.4}Fe_2As_2$ single crystals

The optimally doped $Ba_{0.6}K_{0.4}Fe_2As_2$ single crystals studied here were grown with the self-flux method[1]. The measurements of x-ray diffraction indicate a highly c-axis orientation and good crystallinity of our samples. As shown in Figs.S1**a** and S1**b**, the critical transition temperature of the zero resistivity is about 38.3 K, with a transition width about 0.8 K (between 5% and 95% of the normal state resistivity). A magnetic field of 9 T shifts the upper critical transition (90% of the normal state resistivity) with 1.22 K and zero resistivity temperature with 3 K, which means that this material has extremely high upper critical field. Residual resistive ratio [$RRR \equiv \rho(T=300 \text{ K})/ \rho(T=40 \text{ K})$] taken from Fig. S1**b** is about 10.3. The DC magnetization measurement shown in Fig. S1**c** also indicates a very sharp transition. As shown in Fig. S1**d**, the residual specific heat coefficient obtained by extrapolating the curve $C/T$ to zero temperature is about 2.6 mJ/K$^2$·mol. Such a small residual specific heat coefficient indicates a very small nonsuperconducting volume, or weak impurity-induced quasiparticle excitations in the $Ba_{0.6}K_{0.4}Fe_2As_2$ single crystal studied here.

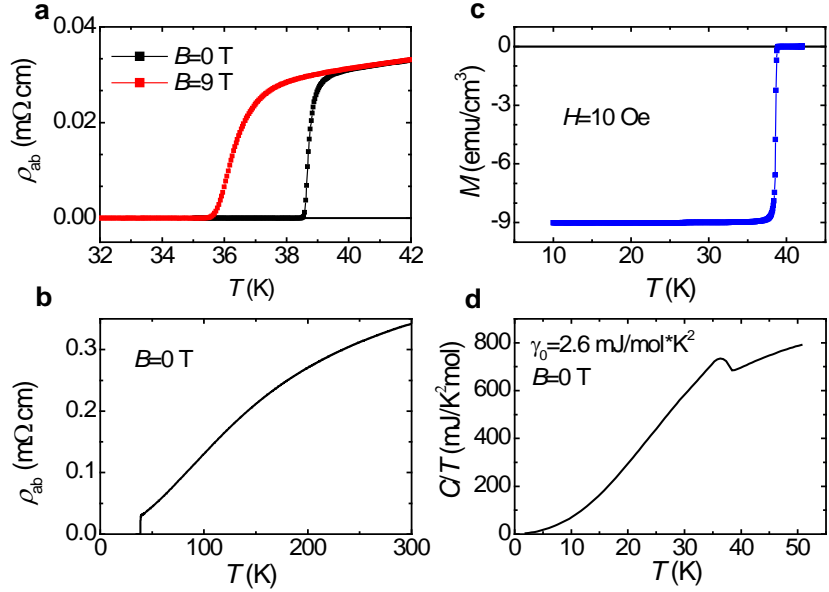

**Figure S1 | Physical characterization of the $Ba_{0.6}K_{0.4}Fe_2As_2$ single crystal. a**, Temperature dependence of resistivity of a $Ba_{0.6}K_{0.4}Fe_2As_2$ single crystal at the temperature near the transition at zero field and 9 T. **b,** Resistivity data versus temperature extended to the room temperature at zero field. **c**, DC magnetization measured at the magnetic field of 10 Oe after zero-field-cooling. **d**, Temperature dependence of specific heat.

## II. STS measurements and fitting

The scanning tunneling spectra (STS) were measured with an ultrahigh vacuum, low temperature and high magnetic field scanning probe microscope USM-1300 (Unisoku Co., Ltd.). The samples were cleaved at room-temperature in an ultra-high vacuum with a base pressure about $1 \times 10^{-10}$ torr. In all STM/STS measurements, Pt/Ir tips were used. To lower down the noise of the differential conductance spectra, a lock-in technique with an ac modulation of 0.1 mV at 987.5 Hz was used.

The tunneling spectra don't exhibit multi-gap feature, i.e., there are no additional superconducting coherent peaks or shoulders at low energy. Considering the multi-gap nature in this material, we first used simplified two-anisotropic-gap model to fit the normalized spectra. In this simplified two-gap model to fit the normalized spectra, the tunneling current is constructed as follows: $G = p dI_L/dV + (1-p) dI_s/dV$, where $I_{L(S)}(V)$ is the tunneling current contributed by the larger (smaller) gap $\Delta_{L(S)}(\theta)$ and $p$ is the related spectral weight. $I_{L(S)}(V)$ is given by

$$I_{L(S)}(V) = \frac{1}{2\pi}\int_{-\infty}^{+\infty}d\varepsilon\int_0^{2\pi}d\theta[f(\varepsilon)-f(\varepsilon+eV)]\cdot\mathrm{Re}\left(\frac{\varepsilon+eV+i\Gamma_{L(S)}}{\sqrt{(\varepsilon+eV+i\Gamma_{L(S)})^2-\Delta_{L(S)}^2(\theta)}}\right), \quad (S1)$$

here, $f(\varepsilon)$ is the Fermi function, and $\Gamma_{L(S)}$ is the scattering factor of the large (small) gap [2]. We take the anisotropic nodeless gap function $\Delta_{L(S)}(\theta) = \Delta_{L(S)}^0(x\cos 2\theta + 1 - x)$, in which x determines the anisotropy of gaps. In all process of fitting, p and x equal to 0.8 and 0.3, respectively. The fitting parameters for the normalized spectrum shown in Fig. **2b** are listed in Table S1.

Table S1 | Fitting parameters for normalized spectrum at different temperatures.

| Temperature(K) | $\Delta_L$ (meV) | $\Delta_S$ (meV) | $\Gamma_L$ (meV) | $\Gamma_S$ (meV) |
|---|---|---|---|---|
| 1.7 | 7.3 | 3.1 | 0.9 | 1 |
| 5 | 7.4 | 2.9 | 1 | 1 |
| 10 | 7.2 | 2.8 | 1 | 1 |
| 15 | 6.8 | 2.7 | 0.75 | 0.8 |
| 20 | 5.8 | 2.4 | 1.1 | 1 |
| 25 | 4.2 | 2.1 | 1.1 | 1 |
| 30 | 3.6 | 1.9 | 1.1 | 1 |
| 35 | 1.2 | 0.5 | 1.1 | 1 |

A single anisotropic gap model also works in fitting the experimental data with a larger scattering factor. As shown in Fig. **S2**, both of the two fitting results can capture the main superconducting feature of the experimental curve at 1.7 K, which reveals that the small gap contributes a tiny (even negligible) spectral weight in this measurement. Since the two fitting processes give quite similar results, we choose the fitting curve with two-gap to extract the varying of area of superconducting coherent peaks as well as the bosonic mode feature.

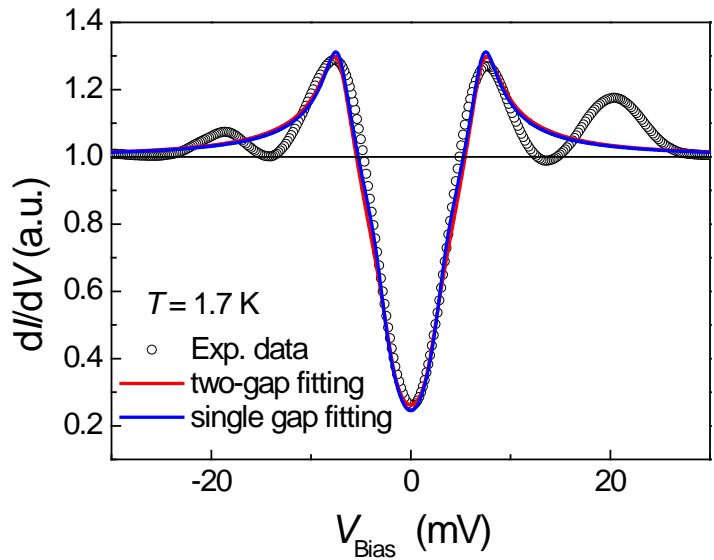

**Figure S2 | Experimental d$I$/d$V$ curve and the fitting of the Ba$_{0.6}$K$_{0.4}$Fe$_2$As$_2$ single crystal at 1.7**

**K.** The symbols represent the experimental STS at 1.7 K normalized by the one measured in normal state (at 40 K), the solid lines are the theoretical fittings to the data by the Dynes model with one single or two anisotropic gaps. The single-band fitting results the same large gap values of 7.3meV as the two-band fitting, while the former requires larger scattering factor.

**III. Simulation based on the Eliashberg theory**

In order to get a qualitative understanding of the bosonic mode, and especially for the asymmetry of the mode features with respect to the Fermi level, we performed self-consistent microscopic calculations using the Eliashberg formalism, assuming that the bosonic mode is the pairing glue. We use a four-band model, the two electron-like bands contribute Fermi pockets centered at (π, 0) and (0, π), while the two hole-like bands contribute pockets at (0, 0) and/or (π, π). The simulated results are presented in Fig.S3 for all the four bands and the summation of them. The electron-boson coupling is greatly appreciated by the scattering between the electron and hole pockets. Because of hole doping in the sample, the bottoms of the electron-like bands (at about -50 meV) are very close to the Fermi level. Considering the scale of the mode energy (~14.5 meV), the conventional wide-band limit does not apply to the shallow electron-like bands. Instead, we model the bare density of states on these bands as $\rho_{e_1,e_2}(\varepsilon) = A_e\sqrt{\varepsilon - \varepsilon_{\min}}$ ($\varepsilon \geq \varepsilon_{\min}$) where $A_e$ is a constant and $\varepsilon_{\min}$ is the energy at the band bottom. (Notice that the bands $e_1$ and $e_2$ are degenerate in energy.) In contrast, the tops of the hole bands are above and relatively more distant to the Fermi level, so we model the bare density of states on such bands ($h_1$ and $h_2$) as constants, $\rho_{h_1,h_2}(\varepsilon) = A_h$. For simplicity but without change of qualitative physics, we ignore the difference in the density of states in the hole bands. Both $\rho_{e_i}$ and $\rho_{h_i}$ are cutoff for $|\varepsilon| > \Lambda$, the energy scale below which well-defined spin-exchange interactions are expected to emerge because of renormalization from quasi-particles at higher energy scales. In our calculation we use $\varepsilon_{\min} = -56 meV$ and $\Lambda \sim 300 meV$. We also tune the parameters $A_{e,h}$ so that the bare density of states $\rho_{e_i}(0) \sim \rho_{h_i}(0)$ for definiteness.

We describe the coupling between the electrons and the bosonic mode by a matrix coupling constant $g_{ab}$, where $a$ and $b$ runs from $e_1$, $e_2$, $h_1$, $h_2$. There are in fact only two independent elements, since because of the scattering vector (π, 0) of the bosonic mode (the dominant spin fluctuations), we may assume $g_{e_i e_j} = g_{h_i h_j} = 0$ and $g_{e_i h_j} = g_{h_j e_i} = g_j$. On the other hand the two hole pockets may have different sizes/shapes, thus they are nested to the electron pockets to different extents. Thus in general $g_1 \neq g_2$.

We assume a sharp Lorentzian density of states $B(\nu)$ for the bosonic mode, with a central peak at $\Omega = 14.5$ meV and a half-width of the order of 0.7 meV (see the inset of Fig. S3). To a good approximation we ignore the intra-pocket structure, so that $\sqrt{\rho_\alpha(0)\rho_\beta(0)}g_{\alpha\beta}^2 B(\nu)$ is the analogue of $\alpha^2(\nu)F(\nu)$ in the usual phonon mechanism. The effective coupling constant is defined as

$$\lambda_{ab} = \sqrt{\rho_a(0)\rho_b(0)}g_{ab}^2 \int_0^\infty d\nu B(\nu)/\nu. \tag{S2}$$

Since we retain the energy dependence in $\rho_a(\varepsilon)$, such quantities must combine $\varepsilon_{\min}$ defined above to determine the superconducting properties. In our calculation we use $\lambda_{e_i h_1} = \lambda_{h_1 e_i} = 0.75$ and $\lambda_{e_i h_2} = \lambda_{h_2 e_i} = 0.3$. The difference of the two sets of coupling constants follows from the consideration for $g_{ab}$ mentioned above.

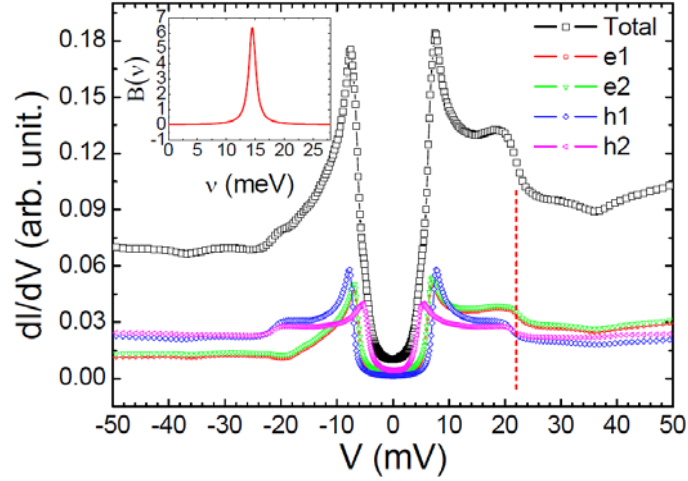

**Figure S3 | Theoretical simulation on the STS based on the Eliashberg theory.** The main panel shows the simulated STS curves in a four-band model using the Eliashberg theory. The mode feature is quite clear at the voltage of about 22 mV, as marked by a vertical dashed-line. Due to the band edge effect, the related features are prominent in the positive-bias side, but very weak in the negative-bias side. The inset is the assumed bosonic distribution function for the simulation, which takes a Lorentzian function and centers at 14.5 meV. Since the two electron bands are highly degenerate, the band $e_2$ has been shifted 0.001 upward for illustration.

The self-consistent Eliashberg equations for the Matsubara Green's function in the Nambu space are given by,

$$G_a^{-1}(\varepsilon, i\omega_n) = i\omega_n \tau_0 - \varepsilon \tau_3 - \Sigma_a(i\omega_n), \quad (S3)$$

$$\Sigma_a(i\omega_n) = \sum_b g_{ab}^2 \int d\varepsilon \rho_b(\varepsilon) T \sum_{\omega_m} G_b(\varepsilon, i\omega_m) \int_0^\infty d\nu B(\nu) \frac{2\nu}{\nu^2 + (\omega_n - \omega_m)^2}, \quad (S4)$$

where $\tau_0$ is the identity matrix, $\tau_3$ is the third Pauli matrix, $T$ is the temperature, and $\omega_n$ is the fermion Matsubara frequency. Notice that we have assumed that the bosonic mode is magnetic in nature, so that the coupling matrix between the fermions and the bosons in the Nambu space is an identity matrix. The information of pairing is contained in the $\tau_{1,2}$ components implicit in the matrix self-energy function $\Sigma_a$. In practice these equations are transformed to real-frequency axis. Once self-consistency is achieved, the renormalized band-resolved density of states is given by

$$N_a(\omega) = -\frac{1}{\pi}\int d\varepsilon\, \rho_a(\varepsilon)\,\mathrm{Im}\,G_a^{11}(\varepsilon, i\omega_n \to \omega + i0^+), \quad (S5)$$

where $G^{11}$ denotes the (11)-component of the matrix Green's function. In the lack of a detailed knowledge of the tunneling matrix element, the tunneling conductivity is simply assumed to be proportional to the superposition of the band-resolved ones, $N(\omega) = \sum_a N_a(\omega)$. In obtaining the results shown in Fig. S3 we used a temperature $T=1.4$ K, and have assumed an ambient elastic scattering rate $\gamma \sim 1.4$ meV in the normal state Green's function before switching on the electron-boson coupling. We emphasize that the above theory automatically produces anti-phase pairing gap functions on the hole and electron pockets, even though these functions are complex and depends on real frequency. On the other hand, the different couplings associated with the hole bands are very effective to produce the two gaps that differ by a factor of 2 (as experimental data reveal). Otherwise a factor of 4 or larger difference in the bare density of states has to be used to achieve the desired gaps. A comparison between the simulated result and the experimental data of a randomly selected STS curve is shown in Fig. **S4**. As we mentioned in the main text, the background of the experimental STS curve varies spatially due to the polar surface, or the local disorder effect. In this sense the simulated result captures the main feature of the data: the asymmetry of the spectrum and the feature of the bosonic mode.

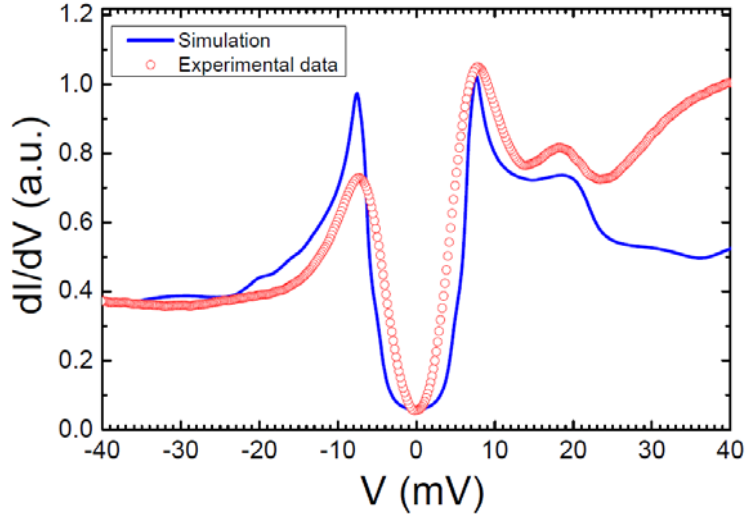

**Figure S4 | Comparison between the simulated result and the experimental raw data of a randomly selected STS curve.** The symbols represent the experimental data, the solid line shows the simulated STS curve by assuming two electron pockets and two hole pockets based on the Eliashberg theory.

**Reference**
[1] Luo, H. Q., Wang, Z. S., Yang, H., Cheng, P., Zhu, X. Y., & Wen, H. H. Growth and characterization of $A_{1-x}K_xFe_2As_2$ (A = Ba, Sr) single crystals with x=0-0.4. *Supercond. Sci. Technol.* **21**, 125014 (2008).
[2] Dynes, R. C., Garno, J. P., Hertel, G. B. & Orlando, T. P. Tunneling study of superconductivity near the metal-insulator transition. *Phys. Rev. Lett.* **53**, 2437 (1984).